\documentclass[11pt,onecolumn]{article}

\pdfoutput=1

\usepackage[margin=3cm]{geometry}

\usepackage[T1]{fontenc}
\usepackage{graphicx}
\usepackage[utf8]{inputenc}
\usepackage{hyperref}
\usepackage{graphicx}

\usepackage{mathtools}
\usepackage{amsfonts}

\usepackage{url}

\usepackage{authblk}

\usepackage[sort&compress]{natbib}

\title{\textbf{pyPESTO: A modular and scalable tool for parameter estimation for dynamic models}}

\author[1,2,3,*]{Yannik Schälte}
\author[4,*]{Fabian Fröhlich}
\author[1,*]{Paul J.\ Jost}
\author[1,*]{Jakob Vanhoefer}
\author[1]{Dilan Pathirana}
\author[2,3]{Paul Stapor}
\author[2,5]{Polina Lakrisenko}
\author[2,3]{Dantong Wang}
\author[1,2,3]{Elba Raimúndez}
\author[1]{Simon Merkt}
\author[2,3]{Leonard Schmiester}
\author[2,3,6]{Philipp Städter}
\author[1]{Stephan Grein}
\author[1]{Erika Dudkin}
\author[1]{Domagoj Doresic}
\author[2]{Daniel Weindl}
\author[1,2,3,$\dagger$]{Jan Hasenauer}

\affil[1]{Life and Medical Sciences (LIMES) Institute, University of Bonn, Bonn 53113, Germany}
\affil[2]{Computational Health Center, Helmholtz Zentrum München Deutsches Forschungszentrum für Gesundheit und Umwelt (GmbH), Neuherberg 85764, Germany}
\affil[3]{Department of Mathematics, Technische Universität München, Garching 85748, Germany}
\affil[4]{Department of Systems Biology, Harvard Medical School, Boston, MA 02115, USA}
\affil[5]{School of Life Sciences, Technical University of Munich, Freising 85354, Germany}
\affil[6]{Leibniz Institute for Natural Product Research and Infection Biology, Jena 07745, Germany\\}
\affil[*]{\textit{These authors contributed equally}}
\affil[$\dagger$]{\textit{Contact: jan.hasenauer@uni-bonn.de}}

\date{}

\begin{document}

\maketitle

\begin{abstract}
    Mechanistic models are important tools to describe and understand biological processes.
    However, they typically rely on unknown parameters, the estimation of which can be challenging for large and complex systems.
    We present pyPESTO, a modular framework for systematic parameter estimation, with scalable algorithms for optimization and uncertainty quantification.
    While tailored to ordinary differential equation problems, pyPESTO is broadly applicable to black-box parameter estimation problems.
    Besides own implementations, it provides a unified interface to various popular simulation and inference methods.
    pyPESTO is implemented in Python, open-source under a 3-Clause BSD license.
    Code and documentation are available on GitHub (\url{https://github.com/icb-dcm/pypesto}).
\end{abstract}

\section{Introduction}


In many research areas, including computational biology, mathematical models are important tools to study complex systems and understand underlying mechanisms \citep{Kitano2002}.
While there are a variety of formalisms to describe biological systems, ordinary differential equation (ODE) models are popular as they provide natural means to describe and explain dynamic changes after perturbations, common in experimental biology \citep{FroehlichLoo2019}.
However, models usually have unknown parameters that need to be estimated – their value and uncertainty – from observed data \citep{Tarantola2005}.
We present pyPESTO, a Python-based parameter estimation tool that provides various inference approaches in a modular manner via a streamlined pipeline (Figure~\ref{fig:concept}; see the Supplementary Information, Section 1, for a tool comparison).
pyPESTO can be applied to various problem types, including large-scale problems.

\section{Features}

\begin{figure}[t]
    \centering
    \includegraphics[width=0.99\textwidth]{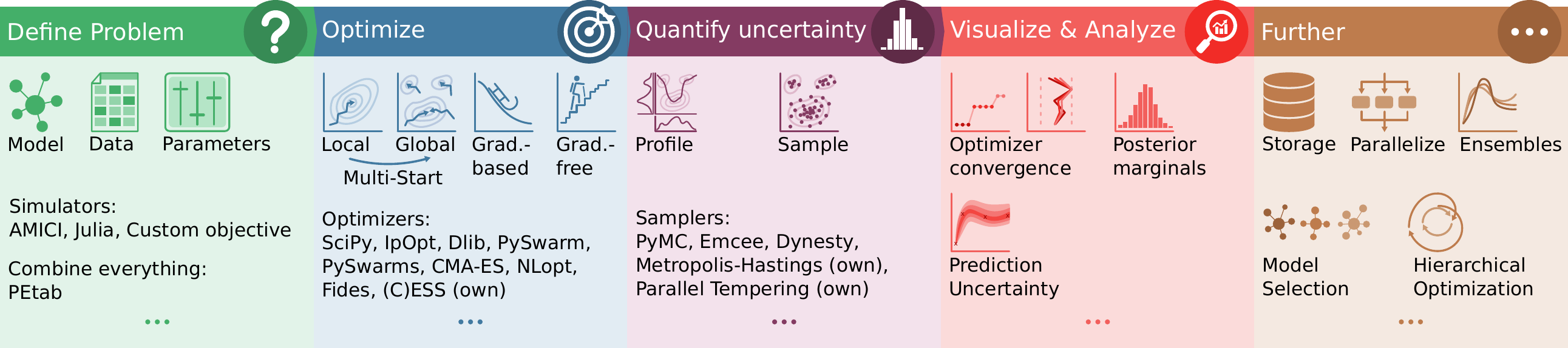}
    \caption{
        Concept figure. pyPESTO covers the full parameter estimation workflow, from problem definition, parameter optimization, uncertainty quantification, to visualization and analysis.
    }
    \label{fig:concept}
\end{figure}


\subsection{Problem definition}

To ease application to biological systems, pyPESTO supports the community standard PEtab for the specification of parameter estimation problems \citep{SchmiesterSch2021}, and interfaces in particular the ODE simulation and sensitivity engine AMICI \citep{FroehlichWei2021}.
Furthermore, pyPESTO allows for user-defined continuous parameter estimation problems given via scalar objective functions or vector-valued residual functions.
This includes log-likelihood and log-posterior based objective functions, where estimated values and uncertainties can be interpreted statistically.
As many inference methods benefit from objective function derivatives, pyPESTO supports user-supplied functions that compute derivatives, but also provides adaptive finite differences.


\subsection{Optimization}

Finding globally optimal parameters as point estimates is a common starting point in parameter estimation. As many problems are non-convex and possess multiple local optima, globalization strategies are necessary.
To this end, pyPESTO provides interfaces to global optimizers as well as a multi-start globalization strategy for local and global optimizers, which performs well for biological problems \citep{RaueSch2013}.
pyPESTO provides a unified interface to local and global optimization libraries such as Ipopt, Dlib, PySwarms, pycma, SciPy, NLopt, Fides (see the Supplementary Information, Section 2.1). An extension to other tools is easily possible.
Moreover, pyPESTO provides a hierarchical approach to efficiently handle relative data and noise parameters \citep{SchmiesterSch2019} and ordinal data \citep{SchmiesterWei2019}.

\subsection{Uncertainty analysis}

For uncertainty analysis, pyPESTO implements the optimization based (frequentist) profile likelihood and (Bayesian) profile posterior approaches using the aforementioned interfaces to different optimization libraries \citep{RaueKre2009}.
Moreover, pyPESTO implements the Bayesian sampling algorithms adaptive Metropolis \citep{HaarioSak2001} and adaptive parallel tempering \citep{MiasojedowMou2013},
and provides a unified interface to the popular sampling tools Emcee, PyMC, and Dynesty (see the Supplementary Information, Section 2.2), supporting in particular gradient-based sampling.


\subsection{Further aspects}

pyPESTO provides various routines to visualize and analyze obtained results. Results can be saved, recovered, and shared in a compact storage format based on HDF5.
Moreover, pyPESTO supports shared-memory parallelization via multi-threading and multi-processing.
Thus, pyPESTO can be deployed flexibly on desktop machines as well as high-performance-computing infrastructure.

\section{Implementation and Availability}

pyPESTO is implemented in Python, open-source under a 3-Clause BSD license.
The code, designed to be modular and extensible, is hosted on \href{https://github.com/icb-dcm/pypesto}{GitHub} and can be installed from \href{https://pypi.org/project/pypesto/}{PyPI}.
Extensive documentation is hosted on \href{https://pypesto.readthedocs.io}{ReadTheDocs}, including numerous notebooks containing tutorials and outlining pyPESTO's functionality.
We ensure correctness during development via unit tests and continuous integration.

\section{Discussion}

pyPESTO has already been used in at least 12 publications, and is continuously being developed by at least 4 core contributors at 3 institutions.
In the future, we plan to implement additional optimization and uncertainty quantification algorithms, to interface pyPESTO with further popular tools, and to extend and further standardize the supported parameter estimation workflows.
We anticipate that pyPESTO will continue to be useful in a variety of computational biology applications and beyond.

\section*{Funding}

This work was supported by the German Research Foundation (DFG) under Germany’s Excellence Strategy (EXC 2047/1 390685813 and EXC 2151 390873048), the SFB 1454 (Metainflammation), TRR333 (BATenergy), AMICI (443187771), the Human Frontier Science Program (LT000259/2019-L1; FF), the German Federal Ministry of Education and Research (BMBF) (01ZX1916A; DW, PL), a Schlegel Professorship at the University of Bonn (JH), and the Joachim Herz Foundation (YS).

\bibliographystyle{natbib}
\bibliography{Database}

\end{document}


\maketitle

\bigskip

\tableofcontents

\newpage

\section{Comparison of simulation and inference tools features}

\begin{table}[!htp]\centering

\caption{Comparison of features of various systems biology simulation and inference tools (line break for readability).}\label{tab: }

\resizebox{16cm}{!}{

\tiny
\begin{tabular}{
| l | l l | l l l l | l l l l | l |
}\hline
Tool &\cellcolor{grey}Supported models &\cellcolor{grey}Modeling/data standards &\multicolumn{4}{l|}{\cellcolor{grey}Optimization} &\multicolumn{4}{l|}{\cellcolor{grey}Uncertainty analysis} &\cellcolor{grey}Parallelization\\
&\cellcolor{grey} &\cellcolor{grey} &\cellcolor{grey}Local &\cellcolor{grey}Global &\cellcolor{grey}Multi-start &\cellcolor{grey}Hierarchical &\multicolumn{2}{l}{\cellcolor{grey}Profiles} &\multicolumn{2}{l|}{\cellcolor{grey}Sampling} &\cellcolor{grey} \\
&\cellcolor{grey} &\cellcolor{grey} &\cellcolor{grey} &\cellcolor{grey} &\cellcolor{grey} &\cellcolor{grey} &\cellcolor{grey}Global &\cellcolor{grey}Local approx. &\cellcolor{grey}MCMC &\cellcolor{grey}PT &\cellcolor{grey} \\\hline
\cellcolor{grey}AMICI &\cellcolor{ok}ODE &\cellcolor{good}SBML, PySB, BNGL, PEtab &\cellcolor{na}N/A &\cellcolor{na}N/A &\cellcolor{na}N/A &\cellcolor{na}N/A &\cellcolor{na}N/A &\cellcolor{na}N/A &\cellcolor{na}N/A &\cellcolor{na}N/A &\cellcolor{good}OpenMP \\
\cellcolor{grey}AMIGO2 &\cellcolor{good}Any &\cellcolor{bad}No &\cellcolor{good}Yes &\cellcolor{good}Yes &\cellcolor{good}Yes &\cellcolor{bad}No &\cellcolor{bad}No &\cellcolor{good}Yes &\cellcolor{bad}No &\cellcolor{bad}No &\cellcolor{bad}No  \\
\cellcolor{grey}Copasi &\cellcolor{good}ODE, MJP &\cellcolor{good}SBML, PEtab &\cellcolor{good}Yes &\cellcolor{good}Yes &\cellcolor{bad}No &\cellcolor{bad}No &\cellcolor{good}Yes &\cellcolor{good}Yes &\cellcolor{bad}No &\cellcolor{bad}No &\cellcolor{bad}No \\
\cellcolor{grey}data2dynamics &\cellcolor{ok}ODE &\cellcolor{good}PEtab &\cellcolor{good}Yes &\cellcolor{good}Yes &\cellcolor{good}Yes &\cellcolor{bad}No &\cellcolor{good}Yes &\cellcolor{good}Yes &\cellcolor{good}Yes &\cellcolor{bad}No &\cellcolor{good}pthreads, parfor \\
\cellcolor{grey}dmod &\cellcolor{ok}ODE &\cellcolor{good}PEtab &\cellcolor{good}Yes &\cellcolor{good}Yes &\cellcolor{good}Yes &\cellcolor{bad}No &\cellcolor{good}Yes &\cellcolor{bad}No &\cellcolor{bad}No &\cellcolor{bad}No &\cellcolor{good}mclapply, dopar \\
\cellcolor{grey}MEIGO64 &\cellcolor{good}Any &\cellcolor{good}PEtab &\cellcolor{good}Yes &\cellcolor{good}Yes &\cellcolor{good}Yes &\cellcolor{bad}No &\cellcolor{bad}No &\cellcolor{bad}No &\cellcolor{good}Yes &\cellcolor{bad}No &\cellcolor{good}jPar \\
\cellcolor{grey}parPE &\cellcolor{ok}ODE &\cellcolor{good}PEtab &\cellcolor{good}Yes &\cellcolor{bad}No &\cellcolor{good}Yes &\cellcolor{good}Yes &\cellcolor{bad}No &\cellcolor{bad}No &\cellcolor{bad}No &\cellcolor{bad}No &\cellcolor{good}MPI \\
\cellcolor{grey}PEPSSBI &\cellcolor{ok}ODE &\cellcolor{ok}SBML &\cellcolor{good}Yes &\cellcolor{good}Yes &\cellcolor{good}Yes &\cellcolor{bad}No &\cellcolor{bad}No &\cellcolor{bad}No &\cellcolor{bad}No &\cellcolor{bad}No &\cellcolor{good}Various \\
\cellcolor{grey}PESTO &\cellcolor{good}Any; exp.\ ODE &\cellcolor{bad}No &\cellcolor{good}Yes &\cellcolor{good}Yes &\cellcolor{good}Yes &\cellcolor{good}Yes &\cellcolor{good}Yes &\cellcolor{good}Yes &\cellcolor{good}Yes &\cellcolor{good}Yes &\cellcolor{good}parfor \\  
\cellcolor{grey}pyBioNetFit &\cellcolor{ok}ODE &\cellcolor{ok}BNGL, SBML &\cellcolor{good}Yes &\cellcolor{good}Yes &\cellcolor{bad}No &\cellcolor{bad}No &\cellcolor{bad}No &\cellcolor{bad}No &\cellcolor{good}Yes &\cellcolor{good}Yes &\cellcolor{good}Various \\
\cellcolor{grey}Tellurium+SBstoat &\cellcolor{good}ODE, MJP &\cellcolor{ok}SBML &\cellcolor{good}Yes &\cellcolor{good}Yes &\cellcolor{good}Yes &\cellcolor{bad}No &\cellcolor{bad}No &\cellcolor{bad}No &\cellcolor{bad}No &\cellcolor{bad}No &\cellcolor{bad}No \\
\hline
\cellcolor{grey}pyPESTO &\cellcolor{good}Any; esp.\ ODE &\cellcolor{good}PEtab &\cellcolor{good}Yes &\cellcolor{good}Yes &\cellcolor{good}Yes &\cellcolor{good}Yes &\cellcolor{good}Yes &\cellcolor{good}Yes &\cellcolor{good}Yes &\cellcolor{good}Yes &\cellcolor{good}Various \\
\hline
\end{tabular}
}

\end{table}

\begin{table}[!htp]\centering

\resizebox{16cm}{!}{

\scriptsize
\begin{tabular}{
| l | l | l | l | l | l |
}\hline
Tool &\cellcolor{grey}Gradients &\cellcolor{grey}Model simulation &\cellcolor{grey}Language &\cellcolor{grey}Last update &\cellcolor{grey}Link \\
\cellcolor{grey} &\cellcolor{grey} &\cellcolor{grey} &\cellcolor{grey} &\cellcolor{grey} &\cellcolor{grey} \\
\cellcolor{grey} &\cellcolor{grey} &\cellcolor{grey} &\cellcolor{grey} &\cellcolor{grey} &\cellcolor{grey} \\\hline
\cellcolor{grey}AMICI  &\cellcolor{good}FW and AJ &\cellcolor{good}Yes &Python, C++ &2023 &\url{https://amici.readthedocs.io/en/latest/} \\
\cellcolor{grey}AMIGO2 
&\cellcolor{ok}FD &\cellcolor{good}Yes &MATLAB &2021 &\url{https://sites.google.com/site/amigo2toolbox/} \\
\cellcolor{grey}Copasi 
&\cellcolor{ok}FD &\cellcolor{good}Yes &C++ &2023 &\url{https://copasi.org/} \\
\cellcolor{grey}data2dynamics  &\cellcolor{good}FD and FW &\cellcolor{good}Yes &MATLAB &2023 &\url{https://github.com/Data2Dynamics/d2d} \\
\cellcolor{grey}dmod 
&\cellcolor{good}FW &\cellcolor{good}Yes &R &2022 &\url{https://github.com/dkaschek/dMod} \\
\cellcolor{grey}MEIGO64  &\cellcolor{good}Can be passed &\cellcolor{na}N/A &MATLAB &2023 &\url{https://github.com/gingproc-IIM-CSIC/MEIGO64} \\
\cellcolor{grey}parPE 
&\cellcolor{good}FW and AJ &\cellcolor{na}N/A (uses AMICI) &C++ &2023 &\url{https://github.com/ICB-DCM/parPE} \\
\cellcolor{grey}PEPSSBI  &\cellcolor{good}FD and FW &\cellcolor{good}Yes &Java &2020 &\url{https://bitbucket.org/andreadega/systems-biology-compiler} \\
\cellcolor{grey}PESTO  &\cellcolor{good}FD, e.g.\ FW + AJ can be passed &\cellcolor{na}N/A (uses AMICI) &MATLAB &2020 &\url{https://github.com/ICB-DCM/PESTO}\\ 
\cellcolor{grey}pyBioNetFit  &\cellcolor{bad}No &\cellcolor{na}N/A (BioNetGen + libroadrunner interfaces) &Python &2023 &\url{https://bionetfit.nau.edu/} \\
\cellcolor{grey}Tellurium+SBstoat  &\cellcolor{good}FW &\cellcolor{good}Yes &Python &2023 &\url{https://github.com/sys-bio/SBstoat} \\
\hline
\cellcolor{grey}pyPESTO  &\cellcolor{good}FD, e.g.\ FW+AJ can be passed &\cellcolor{na}N/A, general interface to simulators, esp.\ AMICI &Python &2023 &\url{https://pypesto.readthedocs.io/en/latest/} \\
\hline
\end{tabular}
}

\end{table}

\begin{table}[!htp]\centering

\resizebox{15cm}{!}{

\scriptsize

\begin{tabular}{ll|llll}
\textbf{Color Coding} &\textbf{} & &\multicolumn{2}{l}{\textbf{Abbreviations}} & \\
\cellcolor[HTML]{d9ead3}Very good & & &N/A &Not applicable (e.g.\ because the tool just does simulation, or just inference, but not both) \\
\cellcolor[HTML]{fff2cc}Ok & & &MCMC &(Single-chain) Markov-Chain Monte-Carlo \\
\cellcolor[HTML]{f4cccc}Not so good & & &PT &Parallel Tempering \\
& & &ODE &Ordinary differential equation \\
& & &MJP &Discrete-space Markov jump process \\
& & &FD &Finite differences \\
& & &FW &Forward sensitivity calculation \\
& & &AJ &Adjoint sensitivity calculation \\
\end{tabular}

}

\end{table}

This overview was compiled to the best of our knowledge, in May 2023.
pyPESTO originated as a reimplementation of the MATLAB tool PESTO \citep{StaporWei2018}, but now offers a streamlined pipeline with a much broader spectrum of modern methods and interfaced tools.

\section{List of interfaced tools}

\subsection{Optimization}

Currently, pyPESTO interfaces the following optimization tools: Ipopt \citep{WachterBie2006}, Dlib \citep{King2009dlib}, PySwarms \citep{Miranda2018pyswarms}, pycma \citep{HansenYos2022cma}, SciPy \citep{PauliGom2020SciPy}, NLopt \citep{JohnsonSch2021nlopt}, and Fides \citep{FroehlichSor2022fides}.

\subsection{Sampling}

Currently, pyPESTO implements the Bayesian sampling algorithms adaptive Metropolis \citep{HaarioSak2001} and adaptive parallel tempering \citep{VousdenFar2016},
and provides a unified interface to the popular sampling tools Emcee \citep{ForemanMackeyHog2013emcee}, PyMC \citep{SalvatierWie2016pymc3}, and Dynesty \citep{KoposovSpe2022dynesty}.

\bibliographystyle{natbib}
\bibliography{Database}